\documentclass[nonacm, sigconf, balance=false]{acmart}

\usepackage{graphicx,pifont}
\usepackage{adjustbox}
\usepackage{caption}
\usepackage{mathtools}
\usepackage{mathrsfs}
\usepackage{xspace}
\usepackage{url}
\usepackage{tikz}
\usepackage{subfig}
\usepackage{pifont}
\usepackage{enumitem}
\usepackage{booktabs} %
\usepackage{listings}
\usepackage[normalem]{ulem}
\usepackage{multirow}
\usepackage{tabularx}

\newcommand{\MAYBE}[1]{}
\newcommand{\TODMECH}[1]{}

\let\oldding\ding%
\renewcommand{\ding}[2][1]{\scalebox{#1}{\oldding{#2}}}%

\newcommand{\one}{\ding[1.2]{172}\xspace}
\newcommand{\two}{\ding[1.2]{173}\xspace}
\newcommand{\three}{\ding[1.2]{174}\xspace}
\newcommand{\four}{\ding[1.2]{175}\xspace}
\newcommand{\five}{\ding[1.2]{176}\xspace}

\newcommand{\app}{smartphone app\xspace}

\newcommand{\server}{server\xspace}
\newcommand{\client}{client\xspace}

\newcommand{\https}{\texttt{HTTPS}\xspace}
\newcommand{\tls}{\texttt{TLS}\xspace}

\newcommand{\POI}{proof-of-intent\xspace}
\newcommand{\PsOI}{proofs-of-intent\xspace}

\newcommand{\myparagraph}[1]{\myparagraphnodot{#1.}}
\newcommand{\myparagraphnodot}[1]{\vspace{4pt} \noindent {\bfseries #1}\xspace}

\newcommand{\md}{smartphone\xspace}

\newcommand{\atk}[1]{$\mathbf{U_#1}$\xspace}

\newcommand{\B}[1]{$\mathbf{B_#1}$\xspace}

\definecolor{mygreen}{rgb}{0.22, 0.72, 0.21}
\newcommand{\updatelater}[1]{#1\xspace}

\newcommand{\numforms}{\updatelater{100}\xspace}

\newcommand{\sysname}{\textsc{IntegriScreen}\xspace}
\newcommand{\name}{\sysname}

\graphicspath{{images/}}

\begin{document}

\title[\sysname: Visually Supervising Remote User Interactions on Compromised Clients]{\sysname: Visually Supervising Remote User Interactions\\ on Compromised Clients}

\author[I. Sluganovic, E. Ulqinaku, A. Dhar, D. Lain, S. \v{C}apkun and I. Martinovic]{
	{\rm
		Ivo Sluganovic\textsuperscript{*},
		Enis Ulqinaku\textsuperscript{$\dagger$},
		Aritra Dhar\textsuperscript{$\dagger$},
		Daniele Lain\textsuperscript{$\dagger$},
		Srdjan \v{C}apkun\textsuperscript{$\dagger$}
		and Ivan Martinovic\textsuperscript{*}
	}\\
	\textsuperscript{*
	}University of Oxford, United Kingdom\\
	\textsuperscript{$\dagger$}ETH Z\"{u}rich, Switzerland 
}

\begin{abstract}

Remote services and applications that users access via their local clients (laptops or desktops) usually assume that, following a successful user authentication at the beginning of the session, all subsequent communication reflects the user's intent. However, this is not true if the adversary gains control of the client and can therefore manipulate what the user sees and what is sent to the remote server.

To protect the user's communication with the remote server despite a potentially compromised local client, we propose the concept of \emph{continuous visual supervision} by a second device equipped with a camera. Motivated by the rapid increase of the number of incoming devices with front-facing cameras, such as augmented reality headsets and smart home assistants, we build upon the core idea that the user's actual intended input is what is shown on the client's screen, despite what ends up being sent to the remote server. A statically positioned camera enabled device can, therefore, continuously analyze the client's screen to enforce that the client behaves honestly despite potentially being malicious.

We evaluate the present-day feasibility and deployability of this concept by developing a fully functional prototype, running a host of experimental tests on three different mobile devices, and by conducting a user study in which we analyze participants' use of the system during various simulated attacks.
Experimental evaluation indeed confirms the feasibility of the concept of visual supervision, given that the system consistently detects over 98\% of evaluated attacks, while study participants with little instruction detect the remaining attacks with high probability.

\end{abstract}

\maketitle

\section{Introduction}
\label{sec:intro}

\begin{figure}[t]
 \centering
 \includegraphics[trim={0 10.4cm 14.4cm 0},clip,width=0.9\linewidth]{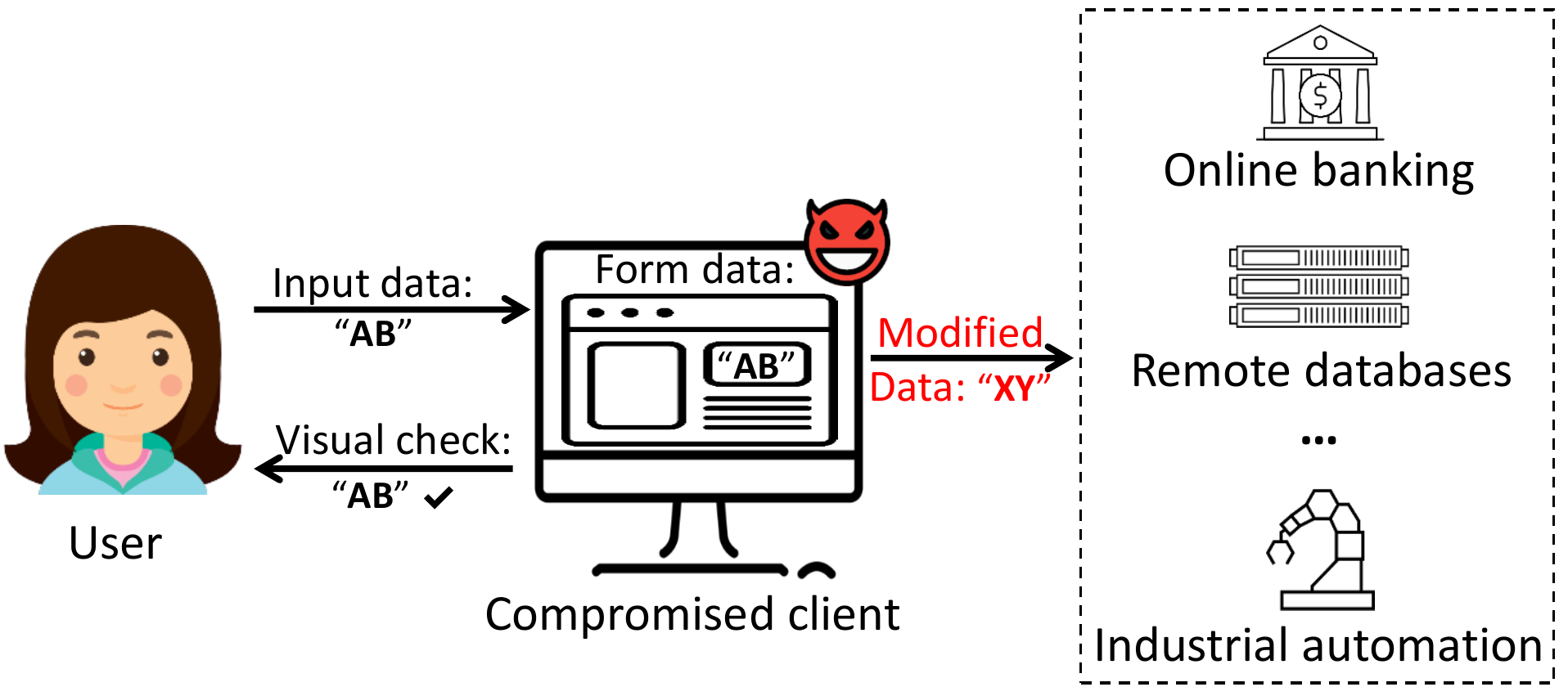}
\caption{\textbf{Motivating scenario.}
 	The user communicates with remote services via the compromised local client.
 	Despite correctly displaying user's input on the screen (``AB''), the adversary arbitrarily sends ``XY'' to the remote service.}
 \label{fig:scenario}
\end{figure}

Users commonly interact with remote online services and applications via web-based interfaces through browsers on their local clients (laptops or desktops). Such services assume that, following a successful user authentication at the beginning of the session, all subsequent communication reflects the user's intent.
However, an attacker that gains control of the client can easily alter the user's actions toward the server after the initial authentication by changing the user input in the background or by changing what the user sees to trick him to provide incorrect input. Figure~\ref{fig:scenario} shows one such example of user input data integrity manipulation where the user intends to input ``AB'' and sees that on the screen, while the remote server actually receives ``XY''. Such compromise of the local client is a realistic assumption due to the large attack surface created by today's complex software and hardware and is often considered during system design of security-sensitive applications~\cite{EUAgencyAssumeHostCompromised}. This is particularly important in scenarios such as online banking~\cite{binsalleeh2010analysis}, accessing private customer data~\cite{uberHack}, or %
configuring safety-critical systems such as industrial~\cite{sadeghi2015security} and medical devices~\cite{ImplantsSecurity}, where a compromised local client can result in a failure, financial loss, or even injury.

As an example, online banking systems assume the possibility of compromise and thus usually mandate that sensitive transaction data are additionally verified by the user.
A common approach is to require that users re-type sensitive fields (such as the beneficiary's account number) into a dedicated hardware token or a smartphone application.
This device serves as a second factor and generates a Transaction Authorization Number (TAN), which the user inputs in their client. %
However, this approach is mostly limited to online banking, duplicates user efforts, and does not scale easily to other applications in which a large amount of sensitive data is inputted by the same operator, e.g., doctors inputting patient data or radiologists configuring medical devices before examinations throughout a typical day.

Another alternative to dedicated hardware for TAN generation is to ask the user to inspect and confirm the transaction details on a smartphone.
This approach does not require dedicated hardware, however, it risks user habituation, as users do not pay sufficient attention to repeated tasks and tend to confirm without reading all details~\cite{userHabituation08}.
Finally, asking users to input the copy of sensitive data on their smartphones would require duplicated effort, increase interaction time, and be error-prone.

In this paper, we are inspired by the general idea of using TANs for data verification; our goal is to extend this concept to support automatic verification of all data that the user inputs into a specific form in their browser.

\myparagraph{Visually extracting user intent}
We build upon the core idea that, during interaction with a large class of networked services and applications, the user's intended input is visually displayed on the client's screen (e.g., in online banking or remote database access).
Despite what a compromised client might attempt to do in the background, users communicate their intention by entering and modifying the values shown on the screen until they are satisfied with what they see, or abort if they are prevented from doing so.

This general idea of supervising user input on an untrusted client to extract user intention has been researched before; however, previous work relied on the assumption that the client is only partially compromised by assuming existence of either a trusted virtual machine~\cite{gyrus}, an operating system~\cite{binder}, or an \emph{attester} application~\cite{nab} that captures the user's input and relays them to the server.

We are motivated by the increase in computer vision capabilities of various camera enabled devices, such as augmented reality headsets~\cite{TimCookAR, HoloLens2}, smart home camera assistants~\cite{fleck2008smart, lenovoSmartHome} and smartphones~\cite{wald2018real, smartphonesCV}.
We thus propose the concept of \emph{visual supervision of user's intent}: by extracting the contents of the client's screen during normal user input, a camera equipped device can infer the values that the legitimate user intends to submit without requiring the duplication of the user's input.
If these extracted values are independently sent to the remote service and compared with each request received from the client, the adversary that controls the client is prevented from either generating arbitrary user input or from modifying the input provided by the user.

As we present in the remainder of this paper, successful implementation of visual supervision of user input requires that multiple technical and research challenges are carefully addressed.
These range from evaluating the technological readiness of existing devices for continuous Optical Character Recognition (OCR) on another computer screen, to preventing concurrent on-screen data manipulation during user input, and detecting and preventing attacks that manipulate users into inputting data under different semantics.
To investigate the design choices and evaluate the technological feasibility of the proposed concept, in this paper we build a functional prototype on an Android smartphone and evaluate it in a series of experiments that confirm the technological feasibility of this approach on existing (and future) devices.

In summary, this paper makes the following main contributions:
\begin{enumerate}[leftmargin=*]
	\item \textbf{System design.}
	We propose and describe \sysname, a system that protects the integrity of the user's input to a remote server by using a device equipped with a camera to visually supervise the user's interaction with an untrusted client and prevents various advanced UI attacks that the adversary might attempt.

	\item \textbf{Prototype \& experimental evaluation.}
	To evaluate the feasibility of the approach on today's smartphones, we build a fully functional prototype of the \sysname system and test it with \updatelater{three} different devices against a range of automated attacks.

	\item \textbf{User study.}
	We run a user study with \updatelater{15} participants, to measure their ability to detect and prevent attacks with \sysname.

\end{enumerate}

\begin{figure}[t]
	\centering
	\includegraphics[trim={0 7cm 17cm 0},clip,width=0.9\linewidth]{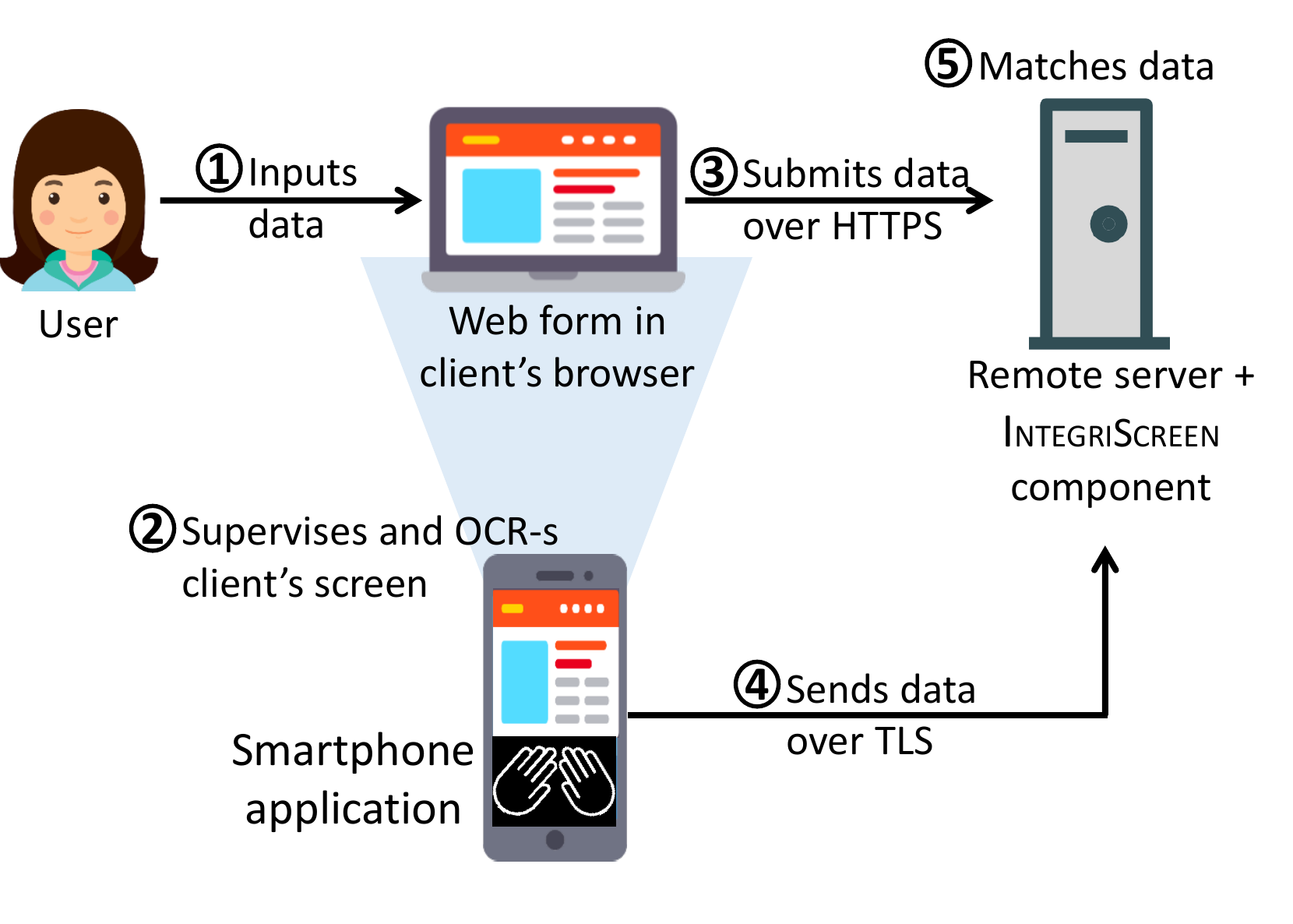}
	\caption{\textbf{\sysname overview.}
	    \one The user inputs data into a web form on an untrusted client, \two while the smartphone's camera captures the interaction.
	    \three Data is submitted both from the client, and \four from the \app.
		\five The server checks authenticity of the received data by comparing the two channels.}
	\label{fig:systemModel}
\end{figure}

\section{Problem Statement}
\label{sec:problemStatement}

We start by motivating the problem of protecting user interactions with the remote web server through an attacker-controlled host; then, we discuss the system and adversary model considered in this paper.

\vspace{0.15cm}
\subsection{Motivation}

Web browsers and web interfaces are nowadays one of the predominant ways to communicate with remote systems.
Many sensitive applications such as e-banking and e-voting, and safety-critical systems such as remote industrial PLCs and medical devices provide web-based interfaces where users provide input parameters.
Typically traditional devices such as desktop and laptop computers running complex software (e.g., operating systems and browsers) are used to access such remote services. Hence, these local clients expose a large attack surface. An attacker-controlled host can modify the input parameters provided by the user and cause fault in the remote services.

\emph{Trusted path}~\cite{x86} provides a secure channel between the user and a trusted service, either on the local system or remote (mediated by a trusted application on the local system).
Possible approaches to establishing trusted paths are \textit{trusted execution environments} (TEEs)~\cite{sgxio} such as Intel SGX, that provide isolation from an untrusted OS, or widespread \textit{transaction confirmation devices}~\cite{filyanov2011uni}, where the user confirms input parameters on an external trusted device.

In this paper, we aim to protect the interactions between the user (who uses a traditional local client) and the remote server without relying on complex solutions such as hypervisors, specialized hardware like TEEs, or systems that introduce significant cognitive loads and risk habituation such as transaction confirmation devices.

\vspace{0.15cm}
\subsection{System and Adversary Model}
\label{sec:problemStatement:systemMode}

Figure~\ref{fig:systemModel} shows an overview of \sysname: the user owns a local client (i.e., a desktop or laptop computer) to provide input to a remote server (service) over the network. We assume that both the client system and the network are fully compromised, while the remote server is a trusted entity. The remote server requires form-based user input, accessible via a web browser running on the client system. As such, all the data whose integrity needs to be protected is presented on the client's screen.
The input can come from a keyboard or a mouse that is connected to the client system. We posit that the user prefers entering input on a computer client, i.e., with a regular mouse and keyboard, rather than using a smartphone with a touchscreen-based keyboard.

The user's interaction with the local client is visually supervised by a trusted application running on a device equipped with a camera, e.g., a smartphone or an augmented reality headset, placed such that it can continuously capture the contents of the client screen during input.
For simplicity, we refer to this device as a smartphone in the rest of the paper -- however, the proposed technique generalizes to any device with network connectivity and a front-facing camera, such as an augmented reality headsets or a smart home assistant device.

\myparagraph{Adversary model}
We assume that the adversary fully controls the client: he adversary can arbitrarily modify the UI elements on the screen, execute keystrokes, mouse clicks, in the device it controls.
Moreover, the adversary controls the network; he can observe, modify or drop any incoming or outgoing network packet from and to the remote server.
However, we assume that the smartphone is not compromised; only the legitimate user can unlock the device and run applications on it.
The remote server is assumed to not be compromised.

The adversary's goal is to achieve that the remote server accepts a request that does not correspond with the legitimate user's intended input.
We consider protecting the privacy of user data and preventing denial of service attacks to be out of scope.

\vspace{0.2cm}
\section{Visual Supervision of User Input}
\label{sec:systemDesign}

We start by providing the general idea of visual supervision of user input; then discuss challenges to be addressed successfully to ensure that interactions with the remote server match user's intentions.

\subsection{Approach Overview}
\label{sec:systemDesign:overallApproach}

We present the overall approach of \sysname in Figure~\ref{fig:systemModel}.
The system consists of 3 main components: (i) the web form and its code, running on an untrusted local client; (ii) the trusted remote \server with a \name web server component; and (iii) the mobile app, running on the smartphone under user's control.

\begin{figure}[t]
    \centering
    \includegraphics[trim={0 1cm 15cm 0},clip,width=0.9\linewidth]{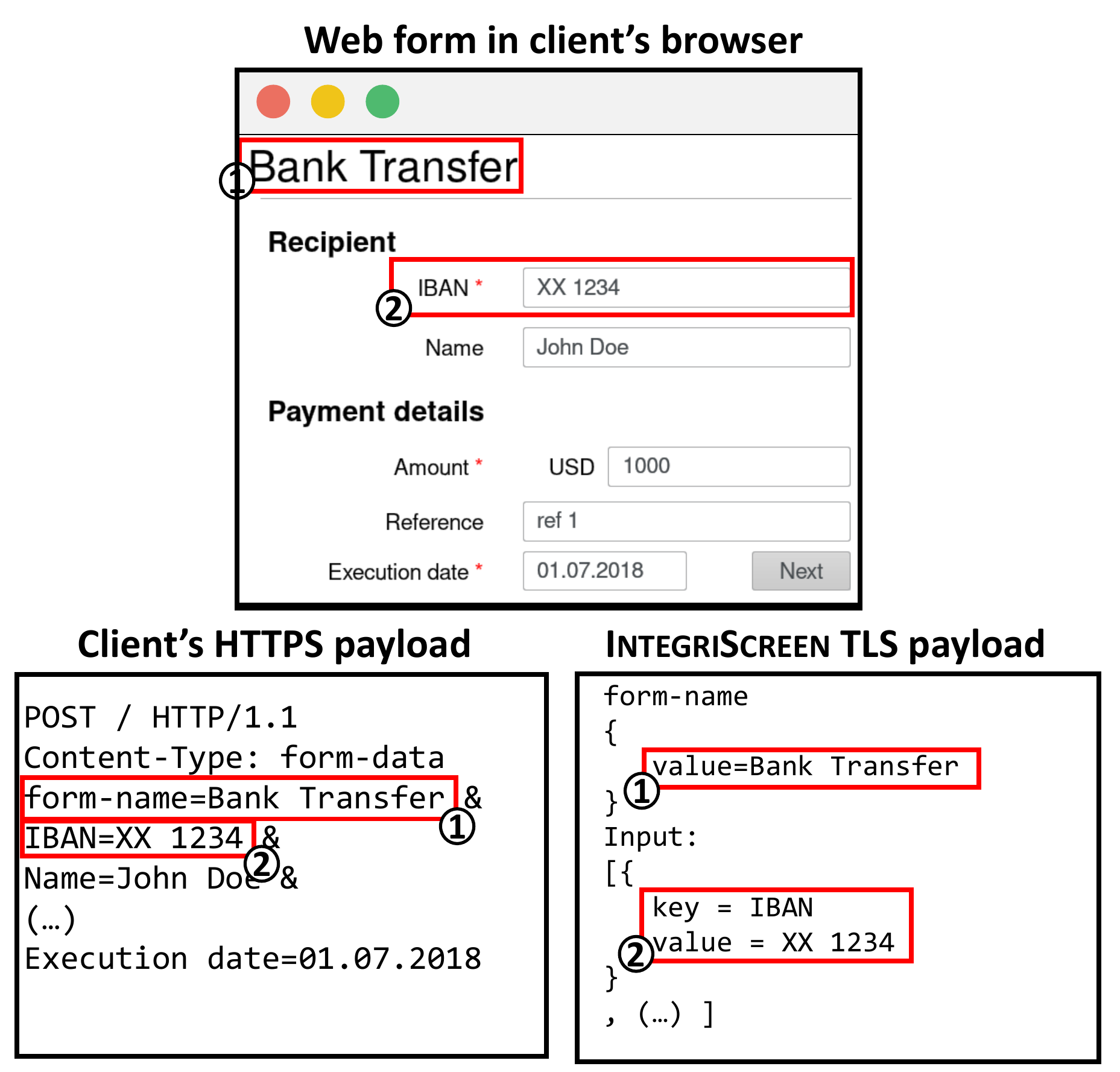}
\caption{\textbf{\name input data matching.}
        The server receives data (e.g., \one and \two) from two channels: (i) the \https channel from the client, and (ii) the TLS channel from the \name mobile app.}
    \label{fig:traceMatching}
    \vspace{0.3cm}
\end{figure}

The general flow of the user's interaction with \sysname is shown in Figure~\ref{fig:systemModel} and described below:
\begin{enumerate} [leftmargin=*]
  \item[\one] The user inputs the data through the form running in the untrusted client's browser.

  \item[\two] The \app extracts the data that is input by the user by performing optical character recognition (OCR) of the client screen in order to generate a visual \POI.

  \item[\three] The browser transfers the user's input over a \https channel to the remote server.

  \item[\four] The \app then transfers the generated \POI to the remote server over a dedicated \tls channel between the smartphone and the remote server.

  \item[\five] Upon receiving data from both the browser and the mobile device, the \name server component matches the data from two inputs, as shown in Figure~\ref{fig:traceMatching}.
  If the two inputs match exactly, the web server accepts the input; otherwise, it rejects it.
\end{enumerate}

The design of \sysname protects against the majority of attacks discussed in Section~\ref{sec:intro}.
Even after compromising the client and obtaining the victim's authentication credentials, the adversary can neither generate new requests, nor covertly modify the data before submitting.
This is prevented by the server, which would reject the requests due to not having received a matching \POI.

\vspace{0.1cm}
\subsection{Remaining Challenges}
\label{sec:systemDesign:challenger}

There remain several other challenges and possible attacks to be addressed to ensure that all received requests truly correspond with the user's intended input, presented in the following.

\myparagraph{Preventing UI manipulation attacks}
With our initial design, the adversary is prevented from modifying the data in transit between the user and the server.
However, he fully controls the client device operations, including loading and presenting the user interface: by modifying elements of the user interface, the adversary can trick the user into entering the \textit{wrong} values -- by changing the semantics of the input.
For example, in case of remote configuration of medical devices, changing a single text label in the user interface (or even only its relative position!) can result in an anesthesiologist entering a value using a wrong measurement unit, e.g. ml instead of dl, and seriously harming the patient due to a tenfold increase in the drug dosage.
These attacks would be successful even if the user re-types his inputs into a special device that generates a TAN (transaction authorization number) code.

As we discuss in Section~\ref{sec:hardenUI}, \sysname prevents such attacks by requiring that the remote server provides a specification of the user interface (UI) and by ensuring that the UI loaded on the client's screen matches its specification during all stages of user input.

\myparagraph{Preventing on-screen data modification attacks}
So far, there was no explicit discussion of the moment at which mobile app captures the data shown of the screen.
Even if the data shown on the client's screen is truthfully sent to the remote server, the integrity of user input is not necessarily guaranteed if data extraction happens only before the form is submitted from the client.
Despite the user entering the intended data, the adversary can subsequently modify the content of the screen in such a way that the user does not notice the change, the \app records it, and the server thus receives the same maliciously modified data from both channels.

For example, in the web form shown in Figure~\ref{fig:traceMatching}, while the user is focused on entering the payment amount or execution date, the adversary could modify the previously input IBAN, without the user noticing.
Furthermore, the adversary could aim to modify the values shown on the screen while the user is absent, not focused on the screen, or a malicious window temporarily overlays the browser form to shift the victim's attention.

As we discuss in Section~\ref{sec:hardenUI}, a crucial step to generate \PsOI\ is real-time, continuous supervision of the screen content, with  specific expectations about the design and behavior of the user interface, and only allowing the mobile app's \POI to be submitted if the data has been generated in accordance to these rules.

\myparagraph{Challenges of visually supervising another device}
Computers still struggle in understanding visual computer interfaces.
For example, even a seemingly straightforward task of detecting computer screens in images is still an open research question~\cite{detectingScreens}.
It is, therefore, an interesting challenge to propose a system that requires \emph{minimal changes to the user interface}, while at the same time achieving security guarantees against the adversarial client system.

Furthermore, despite recent significant improvements in Optical Character Recognition (OCR) based on deep learning~\cite{tesseractOCR}, achieving \emph{consistent continuous detection of textual content} on another device's screen requires several deliberate design choices.
Using OCR libraries naively, e.g. attempting to detect all text shown on a large part of the client's screen, results both in low performance ($<0.5$ fps) and significant parts of text not being detected due to different font sizes and types.

Finally, continuous visual supervision of another device's screen requires that the mobile device is statically positioned so that its camera captures the whole area of interest.
This, however, means that the supervised screen is captured from different positions and different angles and requires \emph{precise estimation and removal of the pose} between the two devices.

\vspace{0.2cm}
\subsection{Design Goals}
Successful user input supervision to extract user intent should:

\begin{enumerate}[leftmargin=*]
	\item Authenticate remote requests that users make through compromised clients, i.e., ensure that the adversary can neither generate nor modify existing remote requests successfully.

	\item Ensure that users are not being manipulated into entering and submitting data that they would not submit in the absence of the adversary.

	\item Require minimal added interaction in the absence of attacks: do not require that users input or explicitly verify any data except on the client device.
\end{enumerate}

\section{\sysname Architecture} \label{sec:hardenUI}

We now describe in detail how \sysname addresses the remaining challenges discussed in the previous section to ensure that all remote requests truly correspond to a legitimate user's intended input.

\subsection{Verifying the Integrity of the User Interface}
\label{sec:systemDesign:webpage}

The precise layout of web forms is as important to protect as the user input: the adversary can alter the semantics of the input by manipulating the UI the user sees.
Sensible changes can be in values, e.g., in the form of Figure~\ref{fig:traceMatching}, changing ``USD'' to another currency to trick into entering a bigger amount, or in element position, e.g., swapping two labels or moving them to trick the user into using the wrong measurement units.
Therefore, it is crucial to assert that the interface the user sees while interacting matches the expectations of the server, both in textual values and in position of the UI elements.

\begin{figure}[t]
	\centering
\includegraphics[width=0.7\linewidth]{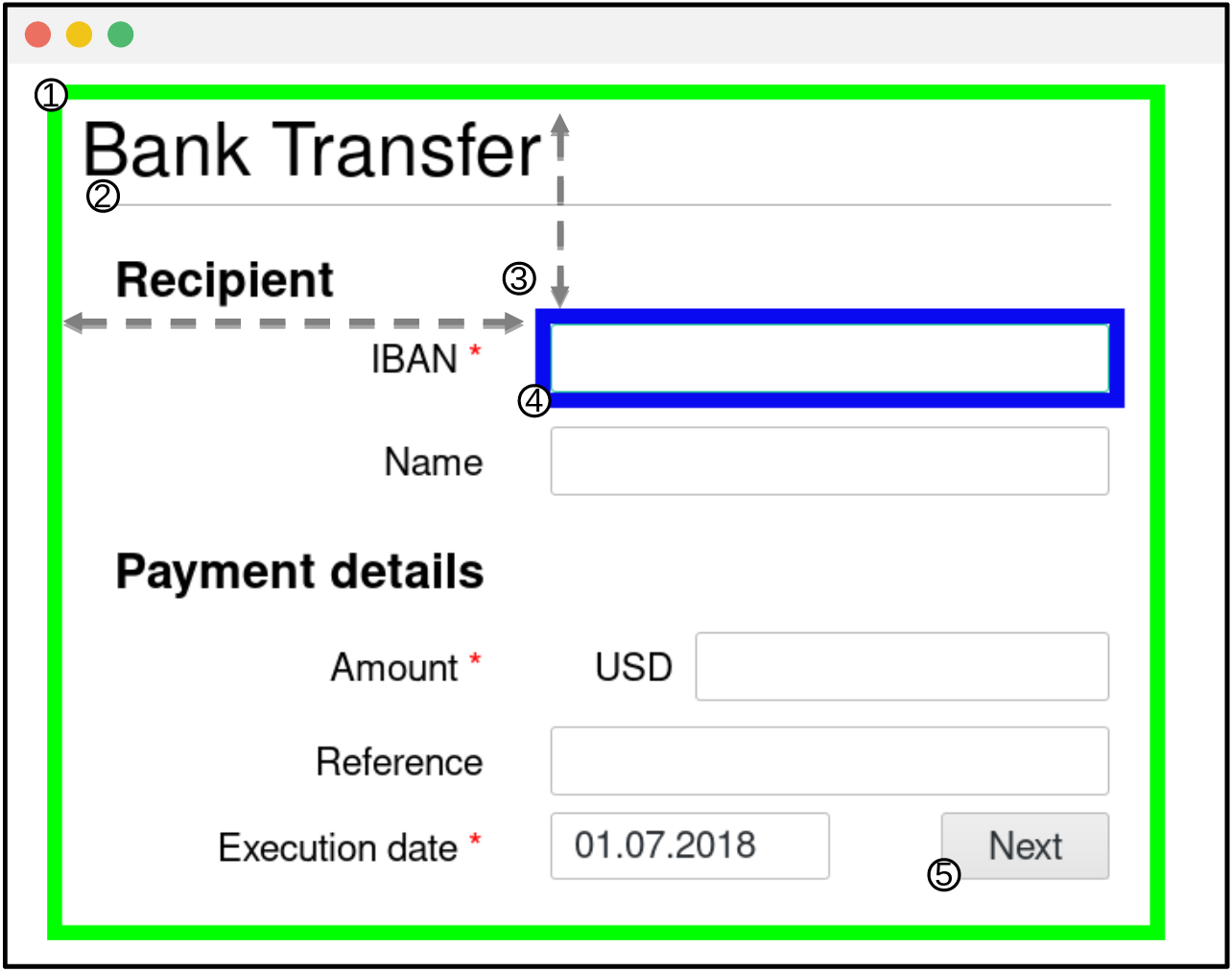}
	\caption{\textbf{\name{} web form features:} \one form boundary, \two form identifier, \three precise positioning of elements, \four the element in focus, \five submit button.
}
	\label{fig:runningExample}
\end{figure}

We now discuss a minimal set of design guidelines that standardize the web form's aspect to harden it against such attacks.
By following these guidelines, a web form can be precisely described in a single JSON form specification file, which the \app obtains from the server, and uses as a guide during supervision of user input.
We show an example of such guidelines in Figure~\ref{fig:runningExample}, and its corresponding form specification in Listing~\ref{code:formSpecification}.

\lstset{
	string=[s]{"}{"},
	stringstyle=\color{blue},
	comment=[s]{:}{,},
	commentstyle=\color{black},
}
\begin{lstlisting}[
float=t,
caption={Form specification example corresponding to the form in Figure~\ref{fig:runningExample}.},
label = code:formSpecification,
basicstyle=\scriptsize\ttfamily]
    "ratio": "1280:960",
    "page_id": "Bank Transfer",
    "elements": [
    {"id": "IBAN_label",
      "type": "label",
      "initialvalue": "IBAN *",
      "x_position": 10,"y_position": 25, "width": 30, "height": 8},
    {"id": "IBAN_value",
      "type": "input",
      "initialvalue": "",
      "x_position": 42, "y_position": 25, "width": 50, "height": 8},
    (...) ]
\end{lstlisting}

\begin{enumerate}[leftmargin=*]
	\item[\one] \textbf{Form Border.}
	\sysname relies on the web form being surrounded by a visible boundary, which serves two roles:
	(i) helping the \app in estimating and removing the pose between the screen and the mobile device for different spatial arrangements; and
	(ii) limiting the area of visual analysis of the screen by excluding other OS/browser UI elements and thus increasing user's privacy.
	For simplicity, the current prototype uses a solid green color to indicate the form border.
	However, this can be fully customized as long as the four corners that indicate the protected area can be detected and tracked by the mobile app~\cite{zhang2002visual}.

	\item[\two] \textbf{Form Title.} \sysname requires that each form has an unique title to download its corresponding specification file.
	While in this paper we assume that the unique form identifier is displayed in the upper left corner, we emphasize its URL could also serve this purpose.
	As we discuss in Section~\ref{sec:securityAnalysis}, the adversary can only cause a denial-of-service by manipulating the form title.

	\item[\three] \textbf{Form Specification.}
	To enable UI verification, \sysname requires that the expected relative positions of all UI elements is known, and a specification of such positions is available to the mobile app after the client loads the web form.
	An example of such specification is given in Listing~\ref{code:formSpecification}: the borders of each UI element are precisely defined relative to the frame border, as are their type (label or input element), and initial values.
	Such specification allows the application to ensure that none of the expected UI elements are missing, modified, or added, thus preventing UI-based manipulation.

	\item[\four] \textbf{Focused Element Border.}
	The input element that is currently being edited (in focus) must be highlighted to indicate the screen area where data changes are allowed to happen.
	While such a visual guide is already implemented on most modern browsers (and we use a blue rectangle for simplicity), \sysname allows for full customization of its design.
	We mandate that while the focused element is changing the remainder of the form (outside of the focus) must remain static.
	Finally, we also restrict how quickly can the focus move away from an element whose value has changed.
	As we discuss in Section~\ref{sec:systemDesign:phone}, this does not limit normal user behavior, while at the same time ensuring stable OCR performance and detection of on-screen modification attacks.

	\item[\five] \textbf{Visible Request Data.}
	Visual supervision requires that the results of entering the sensitive data that comprises the remote request are clearly shown on the client's screen.
	This, for example, allows mouse interaction to change the state of checkboxes or a calendar widget to choose a date, as long as the chosen value is visibly shown afterwards.
	We assume that the form includes a ``Submit'' button that generates the remote request towards the server.

\end{enumerate}

The above requirements are highly extensible, as they do not enforce a specific layout and allow services that implement \name to use their own branding and style.
We further discuss relaxing or modifying some of the requirements in Section~\ref{sec:discussion}.

\begin{figure*}[t]
	\captionsetup[subfigure]{justification=centering}
	\centering
	\null
	\hfill
	\subfloat[The application verifies that all UI elements on the client's screen match their specification, and indicates that users can input data.
		     ] {
		\includegraphics[width=0.5\columnwidth]
		{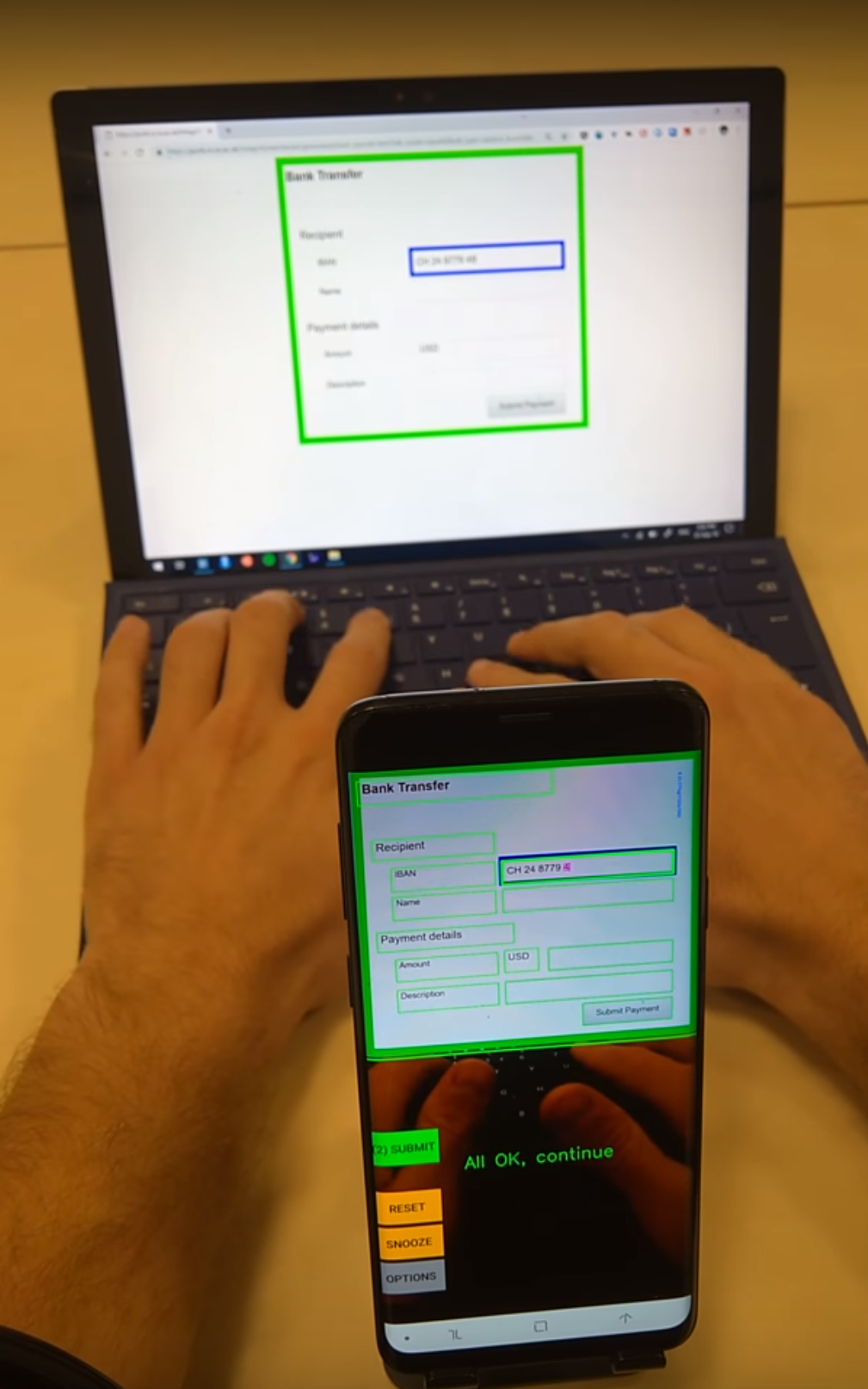}
		\label{fig:userExperience:UIVerificationSuccess}
	}
  \hfill
	\subfloat[Unexpected changes prompt users to stop input, and the mismatch is highlighted.
      The suspicious modification needs to be reverted.
    ] {
		\includegraphics[width=0.515\columnwidth]	{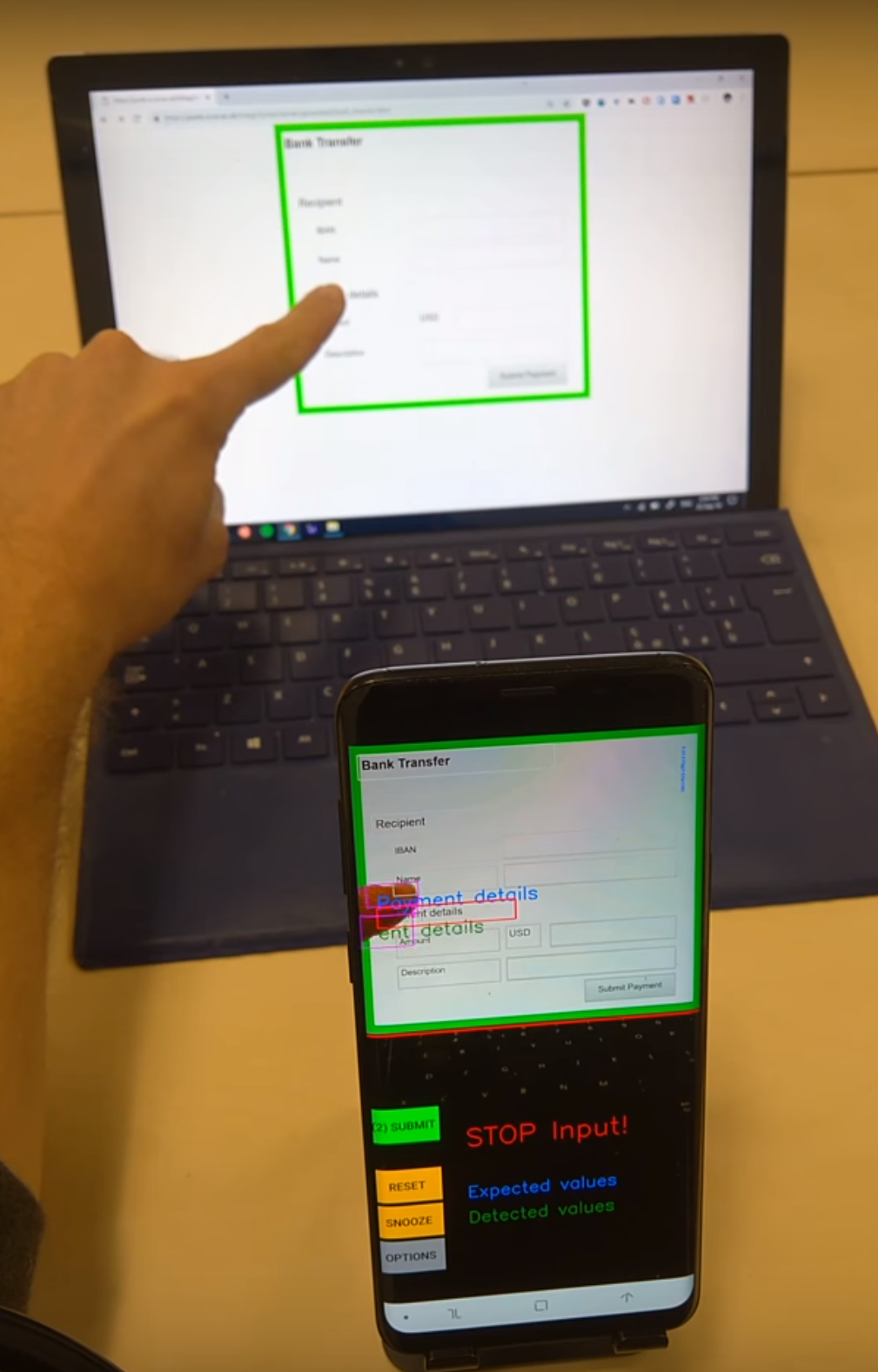}
		\label{fig:userExperience:inputComparison}
	}
	\hfill
	\subfloat[Server comparison mismatch, shown on the smartphone.
		      Client-submitted data is shown in blue; smartphone-submitted data in green.
	] {
		\includegraphics[width=0.545\columnwidth]	{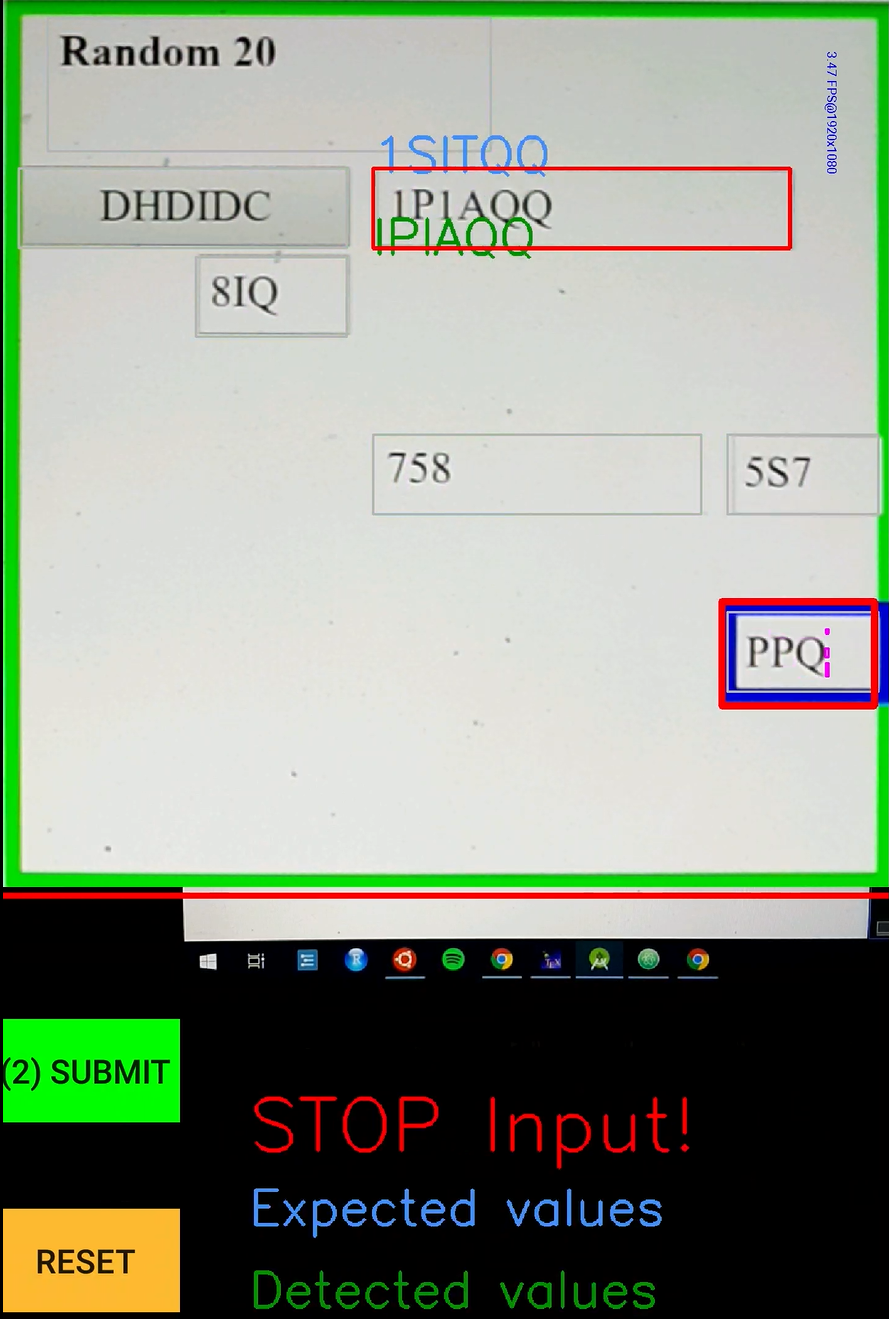}
		\label{fig:userExperience:inputSupervision}
	}
	\hfill
	\null
	\caption{
		User experience of the smartphone application during (a) normal input, (b) unexpected change before submission due to occlusion, and (c) mismatch between client-submitted and smartphone-submitted data.
}
	\label{fig:userExperience}
\end{figure*}

\subsection{Server-side Component}
\label{sec:systemDesign:webserver}

\sysname requires a server-side component that carries out the \emph{input trace matching} between i) the user input from the HTTP payload sent by the browser, and ii) signed input from the \sysname \tls payload. One such example is illustrated in Figure~\ref{fig:traceMatching}. If these two traces mismatch, the \name server-side component notifies the \name mobile app.

\myparagraph{Generating specifications for web forms}
The mobile app requires that the layout of the web form is described in a specification file, which can either be provided by developers of the remote service, or automatically generated by the \sysname server component.
We note that the other direction is also possible: the server component generates the valid HTML web form based on its specification.
We implement this approach for the experimental evaluation in Section~\ref{sec:experimentalEvaluation}, in which we evaluate the performance of the system on hundreds of tests on randomly generated web forms.

\subsection{\sysname Smartphone Application} \label{sec:systemDesign:phone}

The mobile device application provides an independent, out-of-band confirmation of user's intended input despite the client compromise.
This is achieved by verifying that the user interface matches its specification during user interaction, supervising against any on-screen modification attacks, and capturing the data shown on the client's screen to generate and send a respective \POI.

After being loaded by the user, the \sysname mobile app performs the following steps:
\begin{enumerate}[leftmargin=*]
    \item \textbf{Locates} the border of the web form shown on the client's display, realigns the captured video feed to a flat perspective (as shown in Figure~\ref{fig:experimentalSetupRightAngle}), and extracts the form's unique title.
    \item \textbf{Loads} the corresponding UI specification file from the server.
    \item \textbf{[Continuously] verifies that the UI} of the web form shown on the client device matches its specification, i.e., that all UI elements are present and that none have been modified or added.
    \item \textbf{[Continuously] supervises user input}, allowing only the element in focus to change, only when the user is present and active.
	\item \textbf{Submits} the generated \POI to the server.
    \item \textbf{Notifies} the user about the result of server's data comparison: either success (if client and \md -submitted data match) or failure (in case of data mismatch).
    In the latter case, the user is allowed to confirm which of the two versions of submitted data he wants to use (or a combination thereof).
    For additional security guarantees, the user confirms her choice in a hardware protected user interface which is available since Android 9~\cite{androidConfirmation}.
\end{enumerate}

Steps \textbf{(3)} and \textbf{(4)} form the core of the \name mobile app, as they are continuously executed for each frame that the mobile device captures to protect input integrity -- we now describe them in more detail.

\myparagraph{(3) Continuous UI Verification}
For each captured frame, the application verifies that all UI elements conform in position and text values to the expectation.
Labels are never allowed to change, thus always need to match the specification.
Input elements change as a result of user input: when they are modified, the application tracks the latest value input by the user, and verifies that it stays the same afterwards.
Finally, to detect if any UI element has been added by the adversary (which could misguide the user), the application performs text detection on the parts of the frame which should not contain any UI element according to the specification.

In the case of any missing, modified or added UI element, the application warns the user of the potential attack, both visually and by ringing or vibrating.
The application clearly augments its preview of the screen to show which elements are problematic (showing them in red), and prevents any data input until the mismatch is solved.
An example of such warning is shown in Figure~\ref{fig:userExperience:inputSupervision}.

\myparagraph{(4) Continuous Input Supervision}
Besides the layout of the UI, the application also supervises any change on the form to make sure it is a result of intended user interaction:

\begin{enumerate} [label=(\alph*), leftmargin=*]
    \item \textbf{Only the element in focus can change.}
    All other elements must remain the same.
    \item \textbf{Activity detection.}
    If the value of the active element changes, the app must also assure that the user is present. One way to achieve that in our scenario is to detect users hands from the camera feed.
    \item \textbf{Focus changes slowly.}
    If some active element's value changes, the focus should not change to another element in less than $x$~ms after the last edit and in less than $y$~ms since this element first came into focus.
    The server can optionally set the values of $x$ and $y$ in the specification file, otherwise default values are used (in our prototype we set $x=300$~ms and $y=2000$~ms).

\end{enumerate}

Point \textbf{(a)} ensures that the user needs to only pay attention to the value of the currently active element (which they are editing), while all other elements are \emph{protected}.
Point \textbf{(b)} ensures that no element can change while the user is not present and editing the form.
Finally, point \textbf{(c)} serves two goals: (i) ensures that any change can be correctly detected, given frame rate limitations of the mobile app; and (ii) ensures that the adversary can not quickly move the focus to another element and change its value without the user noticing, as this would either last for a minimum of several seconds, or be detected as an attack attempt.
We experimentally evaluate these assumptions in the user study in Section~\ref{sec:experimentalEvaluation} and show an example of input supervision detecting malicious modification in Figure~\ref{fig:userExperience:inputSupervision}.

Furthermore, to prevent the adversary from prematurely submitting the form, the \POI is submitted to the server only after the user explicitly presses the \emph{Submit} button on the mobile device.

\myparagraph{Occlusion and multi-page forms}
Our system design natively supports user interactions in which the form is temporarily occluded (e.g., changing browser tab, minimizing the browser window).
In such cases, UI verification temporarily fails, but will be repeated as soon as the form is again displayed on the client's screen: if the values of all elements are unchanged, user input is allowed again.

Such design also supports multi-page interfaces, as the application will simply store the values of all input elements on each page as the user edits them in arbitrary order.
Clearly, changes to hidden elements are not allowed, and the value is verified every time the element is displayed again (similarly to occlusions).

 \section{Security Analysis} \label{sec:securityAnalysis}

We now informally analyze how \sysname provides authenticated user input under our strong adversary model, from the general setting to more specific attacks.
We assume that the user is not a victim of a targeted attack rather than a wide-spread vulnerability in her OS/applications -- hence it is safe to assume that not all of the user devices are compromised at the same time. Such trust assumption is valid in any system that involves two factors such as OTP.

\myparagraph{Generating arbitrary client requests}
As soon as users authenticate to a remote service, a compromised client would also gain access to their credentials and can subsequently generate arbitrary (authenticated!) requests to the remote server without any user interaction.
However, this is not the case for \sysname, since the remote server requires an authenticated \POI from the \app to accept any request coming from the \client.

\myparagraph{Covertly modifying the client request}
\sysname protects against any adversary that modifies the data exchanged between the \client and the \server without any visible change on the client's screen, because the \POI from the \app and the data received by the \server would mismatch.
This significantly raises the bar for the adversary, who now needs to achieve that the mobile device submits a \POI that exactly matches his malicious data, e.g.,  by tricking the user into submitting data that matches adversary's intentions (i.e., manipulating UI instructions), or by carefully changing the data on the screen so that the modifications are detected by the mobile app, but not by the user.

\subsection{UI manipulation attacks}

If the client is compromised, the adversary fully controls the web form shown to the user on screen and could change the context to manipulate the user to actually input the malicious data himself.

\name prevents such attacks by ensuring that the web form shown to the user directly corresponds to the specification: all labels must show the correct text, all default values of input elements must be present (as their modification could also misguide the user), and no unexpected text is allowed in the rendered web form.

If any of those requirements are not met, the \app clearly shows the offending UI element to the user and does not accept any new input.
We experimentally measure the performance of this UI verification in Section~\ref{ssec:UIVerificationEvaluation} and discuss the potential extension to non-textual elements in Section~\ref{sec:discussion}.

\myparagraph{Modifying the form header}
Since the \app relies on optical recognition of the form title to detect which form specification to load, the adversary can cause the \app to load a form specification that he fully controls by changing the title name.
Note that this only results in a DoS attack: the application uses the same endpoint to load the specification and to submit the \POI; thus, the original server endpoint never receives a matching \POI and the attack is not successful.

\subsection{On-screen data modification}
Besides attempting to manipulate the user into entering incorrect data, the adversary can attempt to directly add, modify, or delete the data shown on the screen. This can happen either before, during, or after user's input. 
The adversary succeeds if such changes are not detected by the user, but registered as legitimate.
\sysname includes several mechanisms to prevent such attacks.

\myparagraph{Not conforming to UI behavior specifications}
The application protects the form semantics by enforcing that the form follows the design guidelines specified in Section~\ref{sec:hardenUI}.
Further, in order to change any element, the adversary must clearly indicate the location of attempted change to the mobile app by showing a blue rectangle around it.
If the web form does not show the blue rectangle to indicate the focused element, no legitimate user input is allowed, resulting only in a DoS.
If multiple rectangles are shown at the same time, or focus changes faster than the form specification allows, the application warns the user.

\myparagraph{Modification during user absence}
The adversary could wait for users to load the form, start the application, and potentially leave their desk without stopping the app.
During this time, the adversary has an opportunity to change any values on the screen, and thus trick the user into submitting modified data.
\sysname, however, prevents such attacks by mandating for user's presence and hand activity during any detected screen changes and raises an alarm if they are absent.
The adversary must thus attempt an attack concurrently with user's interaction with the client.
We discuss other ways to implement this step in Section~\ref{sec:discussion}.

\myparagraph{Concurrent Data Modification}
The application raises an alarm and refuses any change if two elements are shown to be in focus, or if an unfocused element ever changes.
If the user is present and entering data into some element $X$, the adversary is unable to concurrently change any other element $Y$ because $Y$ would also have to be in focus.
We experimentally evaluate the performance against such attacks in Section~\ref{ssec:inputSupervisionEvaluation} and in the user study (Section~\ref{sec:userStudy}).

In case of concurrent modification of the active input element by both the adversary and the user, we assume that the user detects such changes (which are similar to autocorrect not behaving according to user's expectation) and will not move the focus to the next input element until they are satisfied with its content.
We evaluate this assumption in Section~\ref{ssec:userStudyAttackEvaluation} and discuss further measures to reduce this assumption about user behavior in Section~\ref{sec:discussion}.

\myparagraph{Rapid change of focus}
Finally, a potential attack is to change the focused element from $X$ to $Y$ and back to $X$ so that the mobile app detects the change in focus and in value of $Y$, but the user remains unaware (or considers the change to simply be a glitch).
However, if such a change happened too fast, it would raise an alarm due to the limits imposed in the form specification and enforced by the application.
Namely, if the value of a focused element $Y$ is changed, then the form should delay changing the focus to the next element ($X$) for at least 300~ms after $Y$ is changed, and also ensure that the total time that element $Y$ was focused is at least 2~seconds.

The users are thus likely to detect such sudden changes in focus during data input (for at least 2 seconds). We evaluate this assumption as part of our user study in Section~\ref{sec:userStudy}.

 \section{Prototype Implementation} \label{sec:prototypeImplementation}

To evaluate the real-world feasibility of \sysname, we implemented a prototype smartphone and server applications.

\subsection{Mobile application}
The prototype \app is implemented in Android.
We use the Android Text Recognition API~\cite{googleOCR} for optical character detection and recognition.
The rest of core image processing functionality is implemented using OpenCV~\cite{openCV}.

\myparagraph{User experience}
We show our prototype interface in Figure~\ref{fig:userExperience}.
Users have to press a button to start the UI supervision; then, the \app verifies the UI and indicates success by augmenting the camera feed with green rectangles around the UI elements~(Figure~\ref{fig:userExperience:UIVerificationSuccess}).
Users can now proceed with input; once they are finished and submit the data from the client, they press the button ``Submit'' on the \app, to submit the generated \POI to the server.
In case of mismatch, or verification failure, the application shows the offending element in red, together with the two mismatching values: expected, and detected~(Figure~\ref{fig:userExperience:inputComparison}).

\myparagraph{Perspective realignment}
The \md can be positioned on a wide range of angles relative to the \client screen.
Therefore, the first step of the image processing pipeline is to detect the corners of the form boundary (e.g., using hue values), and then use linear perspective realignment~\cite{perspectiveTransform} to crop and reorient the input frame as if it was captured with no angle.

\myparagraph{OCR matching}
Given the inherent limitations of visually detecting whitespace and imperfections in the used OCR libraries, in this prototype we consider two strings equal if they only differ in whitespace and character capitalization.
We note that the proposed system can easily support various levels of matching strictness that would be provided in the form specification.

\myparagraph{User activity detection}
We use the lower part of the camera feed to detect hand activity by filtering out the background of the keyboard, and detecting significant changes between two consecutive frames.

\subsection{Server}

The prototype server is implemented using the Apache Tomcat 9.0 framework~\cite{ApacheTomcat}. %
It serves the web forms requested by the client, accepts the requests and \POI from the client and the mobile app and notifies them of the comparison results.

\begin{table}[t]
  \setlength{\tabcolsep}{10pt}
  \renewcommand{\arraystretch}{1.2}
  \centering
  \footnotesize
\caption{Success rates of UI Verification on 100 randomly generated forms, and overall percentage of correctly detected UI elements.
Forms are displayed for 5 seconds.}
  \begin{tabularx}{\linewidth}{@{}p{0.1\linewidth}lXX@{}}
		& \textbf{Mobile Device}		& \textbf{Forms}	& \textbf{Elements} \\
  	\toprule
  	\multirow{3}{=}{Straight Setup\\ (Fig.~\ref{fig:userExperience:UIVerificationSuccess})}
  		& Samsung Galaxy S9+			&  98\% 			& 99.75\% \\
		& Google Pixel 2XL 			&  93\%				& 97.86\% \\
  		& Samsung Galaxy S6			&  82\% 			& 95.15\% \\
  	\midrule
  	\multirow{2}{=}{Inclined Setup\\ (Fig.~\ref{fig:experimentalSetupRightAngle})}
  		& Samsung G. S9+              & 93\%				& 99.12\% \vspace{.25em}\\
		& Samsung G. S9+ [3 seconds]	& 93\%				& 98.97\% \vspace{.25em}\\
    \bottomrule
  \end{tabularx}
  \label{table:UIVerificationResults}
\end{table}

\section{Experimental Evaluation} \label{sec:experimentalEvaluation}

We now evaluate the guarantees provided by the \sysname system, by running a series of experimental tests against attacks that an adversary might attempt.

\subsection{Preventing UI Manipulation} \label{ssec:UIVerificationEvaluation}

We start by evaluating the performance of UI verification for a set of randomly generated web forms of varying complexity in the absence of attacks, as a baseline for accuracy.
We first compute the performance of several configurations, differing in the used mobile device, relative positioning between the two devices and the total time allowed for the form to be loaded and verified.
We finally analyze in more detail the verification performance of the configuration that we use in the remainder of experimental evaluation.

\myparagraph{Setup}
We use a set of \numforms randomly generated web forms (such as the one in Figure~\ref{fig:userExperience:inputSupervision}), which have a varying complexity of visual elements (between 4 and 9) and types of data as labels and default input values (English words, numerical, and random alphabetical strings).

Each form in the dataset is displayed for exactly 5 seconds.
During this time, the \app automatically detects the form based on its title, loads its specification from the server, and performs UI verification.
Mismatching forms are stored for later analysis.
We evaluate the UI verification performance using three Android devices: Samsung Galaxy S9+, Google Pixel 2XL, and Samsung Galaxy S6.
Additionally, we evaluate the difference in performance when the device is positioned directly in front of the screen and to the right of the keyboard, observing the client screen from an angle.

\myparagraph{Results}
Table~\ref{table:UIVerificationResults} shows the results of verifying the UI for all randomly generated forms. The highest recognition rate was achieved on Samsung Galaxy S9+, which was successful in detecting, loading the specification, and \textbf{verifying the text values of 98\%} of the forms in less than 5 seconds.
This translates to 99.75\% of the UI elements being correctly detected and recognized on the screen.
The other two tested devices also achieved high detection performance: Samsung Galaxy S6 correctly verified 82\% of forms and more than 95\% of UI elements, while these percentages are 93\% and 97.86\% for Google's Pixel 2XL, showing the feasibility of the proposed approach across a range of current mobile devices.
All three devices achieved stable performance, with average processing rates of \updatelater{2.6} (Samsung S6), \updatelater{3.3} (Google 2XL) and \updatelater{4.7} (Samsung S9+) frames per second.

\begin{figure}[t]
	\centering
	\includegraphics[width=0.7\columnwidth]{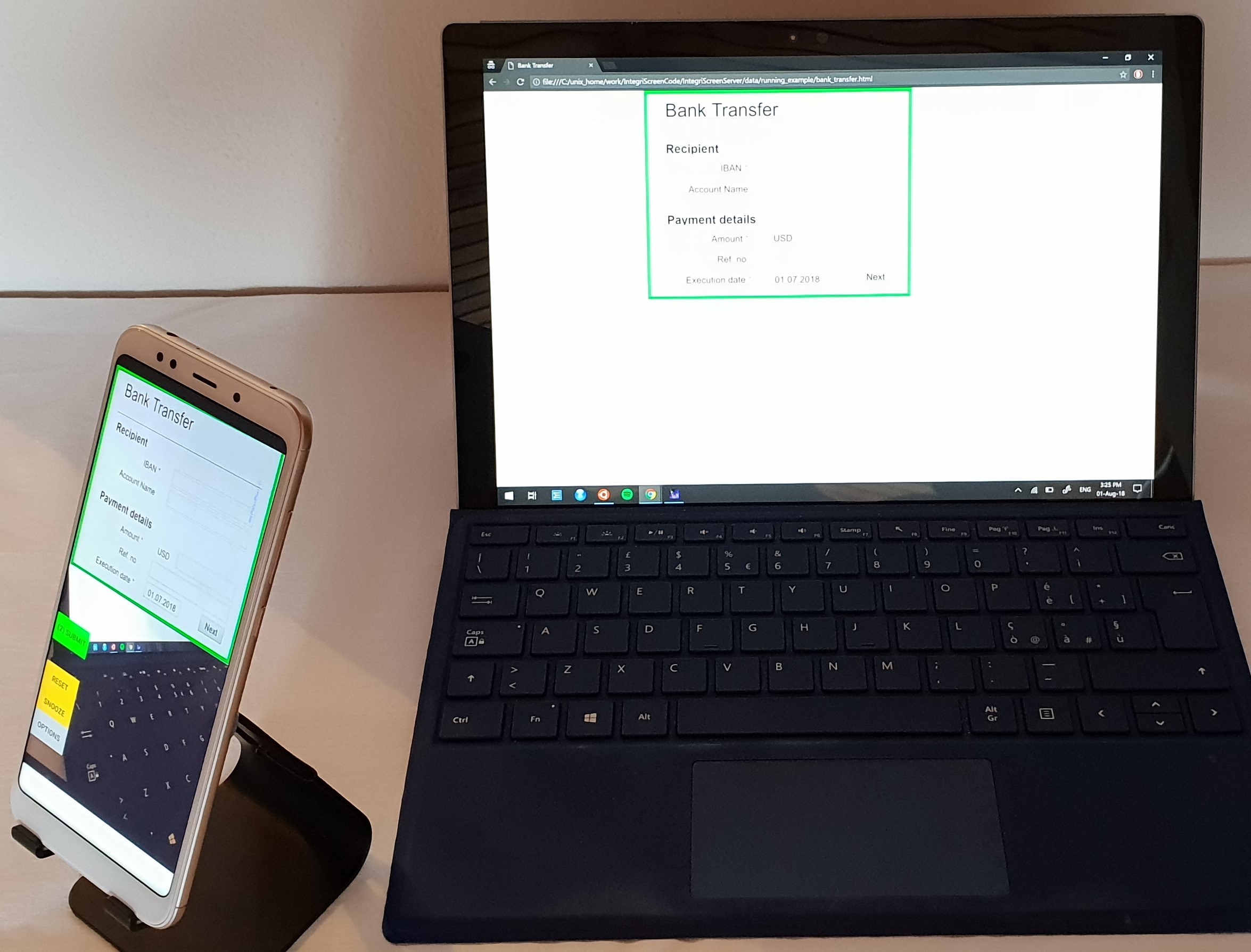}
	\caption{
		\textbf{Alternative experimental setup}.
		\name corrects the angle to a flat perspective before operation.
	} \label{fig:experimentalSetupRightAngle}
\end{figure}

\myparagraph{Positioning and verification time}
Table~\ref{table:UIVerificationResults} also shows the performance of the UI verification procedure for a spatial configuration in which the mobile device is positioned on the right side of the client's keyboard, resulting in a significant angle towards its screen (Figure~\ref{fig:experimentalSetupRightAngle}).
Despite the challenges of precisely realigning and detecting UI elements at a large angle, the evaluation shows that the prototype is successful at \textbf{correctly verifying 93\%} of the forms from the dataset, resulting in an overall per-element detection rate of 99.12\%.

Finally, we measure verification performance when the total time allowed for the application to verify a single form is reduced to 3~seconds, decreasing users' waiting time before they can start input.
The application maintains a high detection rate even for such short intervals: the form verification rate for this short duration remains at a high 93\%, with a per-element detection rate of 98.97\%.

\subsection{Preventing On-Screen Data Modification} \label{ssec:inputSupervisionEvaluation}

We now evaluate the probability that the \sysname app, during user input, detects modification of an element that is not a result of user activity, i.e., one that happens either during page load, or at the time of user input but outside of the focused element.

\myparagraph{Setup}
To achieve testing consistency and allow running a large number of controlled experiments, we simulate user input with Selenium WebDriver~\cite{seleniumWebDriver}, a UI testing framework.
We evaluate potential UI manipulation and on-screen data modification attacks by loading randomly generated forms, and simulating user input by an average \textit{touch} typist (120-200~ms per character)~\cite{pereira2013effect}.

During simulated user input, we run a script that simulates the adversary and replaces three random subsequent characters at a random point in time (Figure~\ref{fig:userExperience:inputSupervision}).
This adversarial change can be one of two different types of attacks that \sysname app must automatically detect and prevent:
\begin{enumerate}[leftmargin=*]
	\item[\B{1}] \textbf{Concurrent modification.} Changing an input element that is not in focus concurrently with the simulated user input\footnote{
A short screen recording of simulating \B{1} attacks can be seen at the following link:
\url{https://tinyurl.com/integriscreen-video}}.
\item[\B{2}] \textbf{Modification before input.} Changing the value of an element while the form is being loaded.
\end{enumerate}

\begin{figure}[t]
	\centering
	\hspace*{-10mm}
	\includegraphics[width=1\columnwidth]{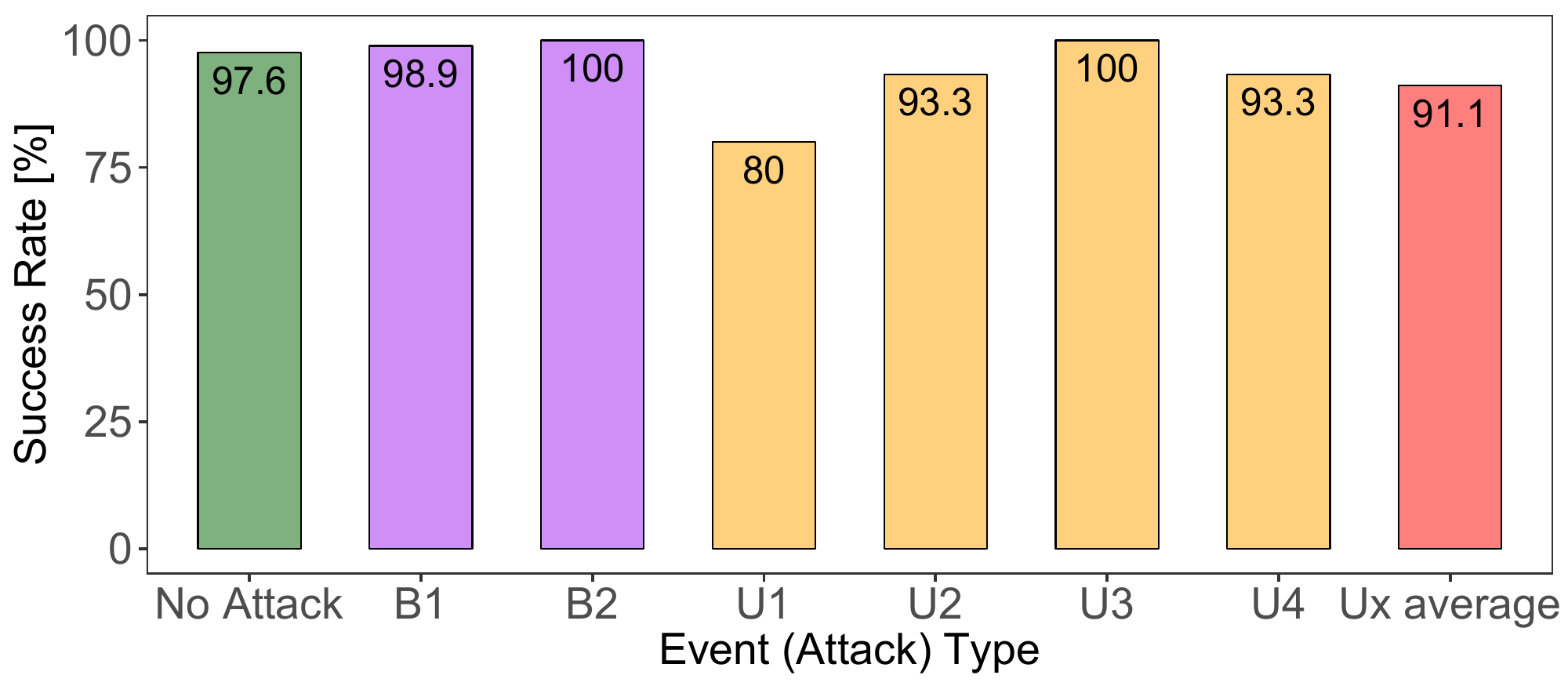}
	\caption{
		\textbf{Overview of the experimental and user study results.}
        In absence of attacks, users successfully completed 97.6\% of form inputs.
        The device detects a large percentage of attacks automatically ($B_1$ and $B_2$), while participants detected more than 90\% of simulated advanced attacks.
	} \label{fig:resultsOverview}
\end{figure}

We consider the attacks prevented if the application stops user input before the form is submitted.

\myparagraph{Results}
As shown in Figure~\ref{fig:resultsOverview}, the application \textbf{successfully detected 98.9\% (91/92) of the simulated attacks of type \B{1}}, preventing the user from submitting data that was maliciously modified during user input.

Further, it was \textbf{successful in detecting 100\% (100/100) of the simulated attacks of type \B{2}}, defending the user from UI manipulation attacks that could have changed the semantics of the form.
Overall, out of 800 UI elements that were loaded (out of which 100 have been modified), the application detected 690 of them to be according to the specification.

In conclusion, our automated evaluation shows that the developed prototype is capable of detecting almost all attacks that are detectable from visual supervision.
In the next section, we evaluate the probability that users detect potential attacks which are hard to distinguish by simply observing the client's screen, and measure if users would successfully understand the potential attacks detected by the implemented system prototype.

\section{Prototype User Study} \label{sec:userStudy}
Finally, in order to evaluate the security guarantees provided by \sysname, we invited participants for a preliminary user study.

\myparagraph{Demographics and instructions}
We recruited a total of \updatelater{15} participants aged between \updatelater{21 and 31}.
The only requirement was a minimum age of 18 years; the test population consists of \updatelater{10} males and \updatelater{5} females.

The participants were first given a short, written description about the experiment\footnote{Given the non-sensitive data that we measured and stored anonymously, our institution did not require an Institutional Review Board approval.
	All study participants, however, received a written description of the study, signed a consent form, and were debriefed after the experiment about the simulated attacks that they experienced.} and the \sysname system (included in the Appendix~\ref{app:userInstructions}).
They were then instructed to fill out 10 forms that represent different online banking transactions by entering the provided test data.
The data for input was given either printed out on paper, or on another screen, as per the participant's choice.
Similarly to Figure~\ref{fig:runningExample}, each transaction required entering the recipient's IBAN (16-24 alphanumeric bank account code), first name and last name, as well as the transaction amount and reference description.

\myparagraph{Experimental setup} \label{ssec:userStudyAttackEvaluation}
While participants were filling out a randomly chosen subset of forms, the client device simulated execution of a total of four different types of attacks on the \sysname system.
The goal of these simulations was to observe user behavior: the likelihood of detecting attacks themselves or correctly understanding the attack detections made by the \sysname prototype.
To observe natural responses and not prime participants to expect potential attacks or specific attack type, they were given no indication that potential attacks might be simulated during the experiment.
The attack types simulated in the user study are:
\begin{enumerate}[leftmargin=*]
	\item[\atk{1}] Modify the value of the focused element during user input.
	\item[\atk{2}] Move the focus to another element, modify its value, and return the focus to the original element.
	The original element is then updated in accordance with user's keypresses that were delayed while focus was shifted.
	\item[\atk{3}] Without moving the focus, modify the value of one of the inactive UI elements while the user is changing another UI element.
	\item[\atk{4}] Modify the value of some element after 3 seconds of user inactivity without moving the focus.
\end{enumerate}
\atk{1} and \atk{2} aim to measure our assumptions about participant's detection of potential attacks which can not be prevented solely by smartphone's visual supervision as the adversary's behavior is not distinguishable from legitimate user's.
\atk{3} and \atk{4} measure the success rate of (i) the mobile app detecting; and (ii) the participants successfully correcting the potential attacks following warnings from \name.

The order of attacks was randomized.
Each attack modified 3 consecutive characters of the same type as the data already on the screen, e.g. digits were replaced with digits.
Attacks were triggered after users made 6 keypresses on a randomly chosen input element (trigger element) for \atk{1}, \atk{2}, and \atk{3}; in case of \atk{4}, the attack was initiated as soon as trigger element modification was followed by 3 seconds without keypresses.
The trigger element and the element that was modified (target element) were randomly chosen.
The only exception was \atk{1}, where both the trigger and target element were always the IBAN field in order to simulate the worst case scenario: the hardest to detect for participants (due to no obvious syntax in IBANs), and the most damaging if successful.

\myparagraph{Results}
We provide the overview of experimental results in Figure~\ref{fig:resultsOverview}. During the experimental user study, a total of \updatelater{56} attack attempts were simulated by the client system (\atk{4} was not always activated).
We consider an attack successful if the participant did not correct the modified data before submitting it to the remote server.
The detection rate of attacks in our study is highly encouraging: users were successful in detecting \updatelater{91.1\%} (\updatelater{51/56}) of simulated attempts, even without receiving any indication or training about potential attack vectors that they should guard against.

The only type of attack that was successful against more than one participant was \atk{1}, in which the adversary concurrently modifies the value of three characters of the IBAN at the same time as participants are copying it from the data sheet.
As a result, their focus inherently switches between the source data and the UI element that they are modifying.
The participants, however, still detected and corrected \updatelater{80\%} (\updatelater{12/15}) of such attempts.
These results are closely in line with previous research, which concluded that participants spent between 70 and 80 percent of time looking at the screen during textual input~\cite{howWeType}.

All remaining attack types were detected in more than 90\% of attempts:
\atk{2}, where users are required to detect that the focus changed to another element without their interaction, was detected in \updatelater{93.3\%} of the attempts.
All users were able to successfully correct \atk{3}, given that the application clearly indicated the element that was modified while not being in focus.
Similarly, only a single user did not correct an instance of \atk{4}, since the value on the screen changed shortly before submitting data from the browser and then from the phone.

In cases where no attack was simulated, \updatelater{3} participants had to reload one of the web forms due to a misrecognition of the OCR engine; no false mismatch was ever reported by the server.
Overall, the participants successfully submitted data for \updatelater{97.6\%} of transactions that they were instructed to perform.

\section{Discussion} \label{sec:discussion}

\myparagraph{Detecting user attention and non-repudiation}
In this paper, the system uses hand movement to detect user's presence and activity.
However, when the mobile device is placed between the client and the user, its front facing camera is well positioned to capture the user's face.
Given the face tracking capabilities available in recent iOS and Android mobile phones, as well as recent advances in mobile camera-based eye tracking~\cite{krafka2016eye}, the system could be extended to precisely track user's attention on the screen and require that gaze is present for certain data modification.
Furthermore, if the mobile device used face recognition to continuously authenticate the user, the system could also provide non-repudiation guarantees.

\myparagraph{Non-textual UI elements}
As the first of the proposed \textit{visual supervision} paradigm, in this paper, we focused on textual input.
However, our approach can be extended to support non-textual UI elements -- as long as their final state is shown on the screen -- such as checkboxes, sliders, or calendar widgets.
Furthermore, while we focused on text extraction, this step could be implemented by a more literal comparison of the client's screen with a screenshot of the web form, as rendered by the server with same aspect ratio and element position.

\myparagraph{Privacy and security of visual supervision}
While continuously recording one's interaction with another electronic device seems intrusive, we note that all processing in \sysname happens on the mobile device.
Therefore, the server only receives a duplicate of the data from the client.
If sending a duplicate of the data is not suitable for any reason, it is straightforward to modify \name to compute and only send a digest (similar to a TAN) to the server.

However, a visual channel gives users clear control over which data the mobile device observes, i.e., only what is shown on the screen at a given moment -- while most OS level applications typically get unrestricted access to the whole system and could violate the users' privacy in the background while keeping them oblivious.

Finally, using only a visual channel between the client and the mobile device reduces the likelihood of a smartphone compromise, since it never directly communicates with the already compromised client.

\section{Related Work} \label{sec:relatedWork}

Previous work on trusted path either relies on a trusted hypervisor that supervises a compromised virtual machine, or on the use of another trusted device that serves as a second factor.

\myparagraph{Trusted hypervisors}
Trusted hypervisors and secure micro-kernels are a possible choice for trusted path. Sel4~\cite{klein2009sel4} is a functional hypervisor that is formally verified and has a kernel size of only $8400$ lines of code. In work done by Zhou et al.~\cite{x86}, the authors proposed a generic trusted path on $x86$ systems in pure hypervisor-based design. Examples of other hypervisor-based works can be found in systems such as Overshadow~\cite{Overshadow}, Virtual ghost~\cite{criswell2014virtual}, Inktag~\cite{hofmann2013inktag}, TrustVisor~\cite{mccune2010trustvisor}, Splitting interfaces~\cite{ta2006splitting}, $SP^3$~\cite{yang2008using}.

Our approach is most similar to Gyrus~\cite{gyrus}, a system that enforces integrity of user-generated network traffic of protected applications by comparing it with the text values displayed on the screen by the untrusted VM.
However, Gyrus requires application-specific logic and does not prevent potential UI manipulation attacks.

Not-A-Bot (NAB)~\cite{nab} ensures that the data received from the client was indeed generated by the user rather than malware by having the server require a proof (generated by a trusted \emph{attester} application) of user's keyboard or mouse activity shortly before each request.
Similarly, BINDER~\cite{binder} focuses on detecting malware break-ins and preventing data exfiltration by implementing a set of rules that correlate user input with outbound connections.
While these approaches are similar to \sysname in ensuring that outgoing requests match user's activity on the client device, our solution differs in that it allows for a fully compromised client.

\myparagraph{Trusted devices}
Assuming a fully compromised client mandates an additional trusted device is used to secure the interaction with the remote server.
Weigold et al. propose ZTIC~\cite{weigold2011}, a device with simple user input and display capabilities, on which users confirm summary details for a banking transaction.
Another approach is taken by Kiljan et al.~\cite{6978928}, who propose a simple \emph{Trusted Entry Pad}, that computes signatures of user-input sensitive values and sends them independently to the server for verification.
However, such approaches either require users to input data, which breaks the normal workflow and duplicates efforts, or require them to confirm transaction details, which leads to habituation and decreased security.
The approach of continuous visual supervision improves on previous work by neither requiring additional input, nor relying on user attentiveness during transaction confirmation.

\section{Conclusion} \label{sec:conclusion}

This work is based on the realization that video capture and processing capabilities of mobile devices have become sufficient for continuous visual analysis of other electronic devices.
In this paper we thus propose the concept of \emph{visual supervision of user input}, that relies on a trusted camera enabled device to force a compromised client to behave honestly by analyzing the client's screen during user input and independently sending the extracted data to the remote server.

We show the feasibility of this approach by developing a fully functional prototype on an Android smartphone, evaluating it with a series of experimental tests, and running a user study in which we measure participants' responses to simulated attacks.
The results confirm the technical feasibility of the proposed concept on existing smartphones; the system prototype automatically prevents simulated attacks in more than 98\% of attempts, while participants in the study detect the majority of the remaining attacks.

Considering the rapid increase in processing power and camera quality of smartphones, but also novel platforms such as augmented reality headsets and smart home assistants, we believe this paper to be an important first step towards deployment of the concept of visual supervision to secure sensitive users' interactions with potentially untrusted devices.

\bibliographystyle{ACM-Reference-Format}
\bibliography{refs}

\appendix
\section*{Appendices}

\begin{figure*}[tp]
	\centering
	\includegraphics[trim={0.6cm 0 0 0.5cm},width=1.05\linewidth]{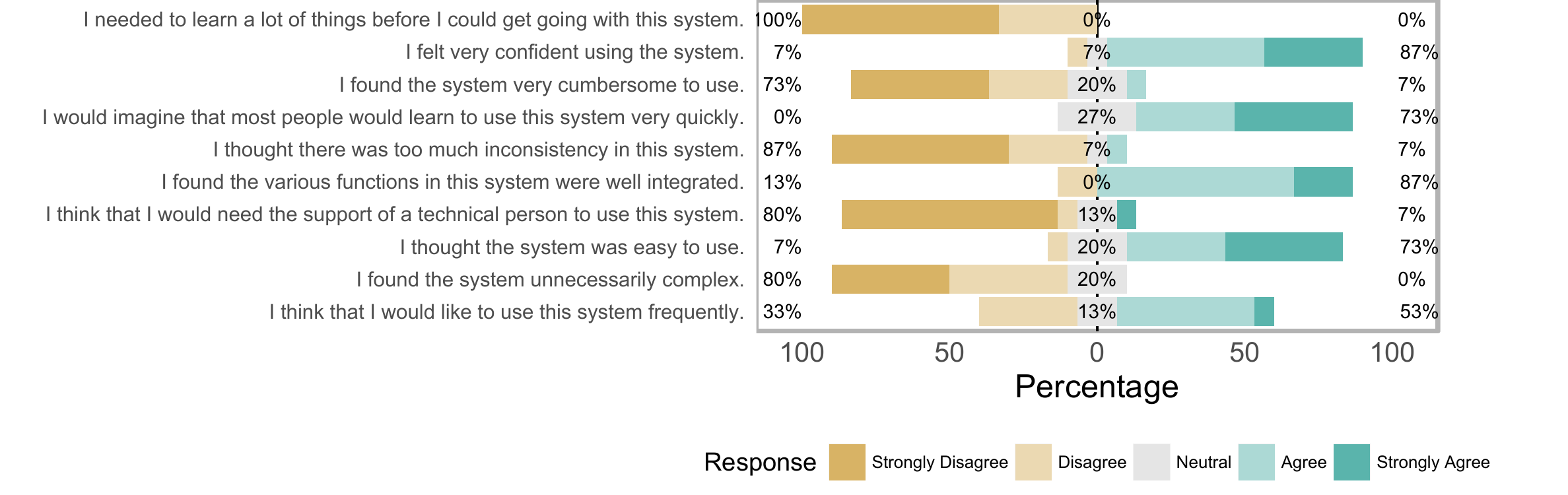}
	\caption{
		\textbf{Participants' responses} to the SUS questionnaire show a high average SUS score of \updatelater{78.3}.
		While some participants felt they would need a help of a technical person (Q4), most generally found the system easy to use (Q3), felt confident using it (Q9), and believed others would learn to use the system very quickly (Q7).
	} \label{fig:SUSResults}
	\vspace{1cm}
\end{figure*}

\section{User Study Instructions} \label{app:userInstructions}

{\sf
"Thank you for accepting to take part in our experimental user study.
In this study, you will simulate several online banking payments (using provided test data).
While you do this, your input will be protected by \sysname, an application that we are currently developing.

\sysname aims to secure users who are sending data to remote servers against the malware that might be present on their computers.
In essence, the app recognizes what you type on the computer screen and sends this data directly to the remote server (a test banking server in this experiment).
By comparing the data stream received from the computer and the stream from the mobile app, the server can verify if the computer is acting honestly.
This system can thus prevent adversaries from, e.g. modifying your transaction details or creating arbitrary transactions."
}

\myparagraphnodot{\sysname usage instructions:}

{ \ttfamily
\begin{enumerate}
	\item Press "START" after loading the form on the computer.

	\item When the app shows "Everything OK", you can input or modify the values on the computer.

	\item If the app detects unexpected behavior on the screen, it will make a sound, show "Stop input!", and indicate the offending element in red rectangle.
	\begin{itemize}
		\item The "Expected" and "Detected" values will be shown in different colors on the screen.
		\item If you believe this is a result of misrecognition, simply correct the values.
		\item Otherwise, abort your input by clicking "RESET"
	\end{itemize}

	\item When you finish with input, click "Submit" on the computer, and then click "SUBMIT" on the mobile device.
\end{enumerate}
}

\section{Usability Questionnaire}
At the end of the experiment, participants were asked to fill out a Simple Usability Score (SUS) questionnaire~\cite{SUS}: a well-accepted tool for evaluating the usability of systems and products~\cite{interpretingSUS}.
The questionnaire consists of 10 Likert scale statements (1 - Strongly disagree to 5 - Strongly agree) -- the overall usability score is computed by adding the scores on odd-numbered questions (e.g. \emph{"Q9: I felt confident using the system"}), and subtracting the scores on even-numbered questions (e.g. \emph{"Q6: I thought there was too much inconsistency in this system"}), and normalizing the value between 0 and 100.

\myparagraph{Results}
The scores of users' perception of the prototype system are shown in Figure~\ref{fig:SUSResults}.
None of the participants thought that they had to learn a lot before they could get going with the system (Q10) or agreed with the statement that the system was unnecessarily complex (Q2).
While some felt they would need the help of a technical person to start using the system (Q4), most generally found the system easy to use (Q3), felt confident using it (Q9), and believed others would learn to use the system very quickly (Q7).

We note that participants in our study evaluated the usability of \sysname only after being required to consecutively input 10 transactions, along with unexpected behavior as a result of attack simulations that some participants perceived as technical glitches of the system rather than malicious behavior.

However, the overall SUS score given by the participants was a high \textbf{\updatelater{78.3}}.
Previous research on interpreting individual SUS scores with a single adjective would thus place the developed prototype between "Good" (mean SUS of 71.4) and "Excellent" (mean SUS of 85.5)~\cite{interpretingSUS}.
The high rates of attack detection and the achieved usability scores of the developed prototype thus allow us to conclude that the experimental evaluation confirms the potential of deploying visual supervision of user input in the future.

\end{document}